\documentclass[11pt,a4paper]{article}
\usepackage{amsmath,amssymb}
\usepackage{jheppubx}
\usepackage{graphicx}

\usepackage[usenames,dvipsnames]{xcolor}

\input epsf
\usepackage{epstopdf}
\usepackage{ upgreek }

\definecolor{darkgreen}{rgb}{0,0.4,0}
\definecolor{darkred}{rgb}{0.4,0,0}
\definecolor{darkblue}{rgb}{0,0,0.4}

\def\be{\begin{equation}}
\def\ee{\end{equation}}
\newcommand{\bea}{\begin{eqnarray}}
\newcommand{\eea}{\end{eqnarray}}

\input epsf
\newcommand{\vev}[1]{\langle #1 \rangle}

\newlength{\extraspace}
\newlength{\extraspaces}
\def\bra#1{{\langle}#1|}
\def\ket#1{|#1\rangle}

\def\vev#1{\langle{#1}\rangle}

\def\II{\relax{I\kern-.10em I}}

\def\IZ{\relax{\rm Z\kern-.34em Z}}
\def\IB{\relax{\rm I\kern-.18em B}}
\def\IC{{\relax\hbox{$\inbar\kern-.3em{\rm C}$}}}
\def\ID{\relax{\rm I\kern-.18em D}}
\def\IE{\relax{\rm I\kern-.18em E}}
\def\IF{\relax{\rm I\kern-.18em F}}
\def\IG{\relax\hbox{$\inbar\kern-.3em{\rm G}$}}
\def\IGa{\relax\hbox{${\rm I}\kern-.18em\Gamma$}}
\def\IH{\relax{\rm I\kern-.18em H}}
\def\II{\relax{\rm I\kern-.18em I}}
\def\IK{\relax{\rm I\kern-.18em K}}
\def\IP{\relax{\rm I\kern-.18em P}}

\def\inbar{\,\vrule height1.5ex width.4pt depth0pt}

\def\IR{\relax{\rm I\kern-.18em R}}

\def\lp10{\ell_p^{10}}
\def\lp11{\ell_p^{11}}
\def\R11{R_{11}}

\def\frac#1#2{{#1 \over #2}}

\newdimen\tableauside\tableauside=1.0ex
\newdimen\tableaurule\tableaurule=0.4pt
\newdimen\tableaustep
\def\phantomhrule#1{\hbox{\vbox to0pt{\hrule height\tableaurule width#1\vss}}}
\def\phantomvrule#1{\vbox{\hbox to0pt{\vrule width\tableaurule height#1\hss}}}
\def\sqr{\vbox{%
  \phantomhrule\tableaustep
  \hbox{\phantomvrule\tableaustep\kern\tableaustep\phantomvrule\tableaustep}%
  \hbox{\vbox{\phantomhrule\tableauside}\kern-\tableaurule}}}
\def\squares#1{\hbox{\count0=#1\noindent\loop\sqr
  \advance\count0 by-1 \ifnum\count0>0\repeat}}
\def\tableau#1{\vcenter{\offinterlineskip
  \tableaustep=\tableauside\advance\tableaustep by-\tableaurule
  \kern\normallineskip\hbox
    {\kern\normallineskip\vbox
      {\gettableau#1 0 }%
     \kern\normallineskip\kern\tableaurule}%
  \kern\normallineskip\kern\tableaurule}}
\def\gettableau#1 {\ifnum#1=0\let\next=\null\else
  \squares{#1}\let\next=\gettableau\fi\next}

 \def\eqnn#1{\xdef #1{(\secsym\the\meqno)}\writedef{#1\leftbracket#1}%
 \global\advance\meqno by1\wrlabeL#1}
 \def\eqna#1{\xdef #1##1{\hbox{$(\secsym\the\meqno##1)$}}
 \writedef{#1\numbersign1\leftbracket#1{\numbersign1}}%
 \global\advance\meqno by1\wrlabeL{#1$\{\}$}}
 \def\eqn#1#2{\xdef #1{(\secsym\the\meqno)}\writedef{#1\leftbracket#1}%
 \global\advance\meqno by1$$#2\eqno#1\eqlabeL#1$$}

\global\newcount\itemno \global\itemno=0

\def\itemaut#1{\global\advance\itemno by1\noindent\item{\the\itemno.}#1}

\def\({\left(}
\def\){\right)}

\def\ii{{\bf i}}

\newcommand{\normord}[1]{\vcentcolon\mathrel{#1}\vcentcolon}
\providecommand{\vcentcolon}{\mathrel{\mathop{:}}}

\usepackage[yyyymmdd,hhmmss]{datetime}
\newif{\ifeq}           
\eqtrue                 
\def\question#1
{
\ifeq
\textcolor{darkblue}{#1}
\fi
}

\nocite{*}

\begin{document}

\title{Nonrelativistic Conformal Field Theories in the Large Charge Sector}
\author{S.M. Kravec \&}
\author{Sridip Pal}  
\affiliation{Department of Physics, 
University of California at San Diego,
La Jolla, CA 92093} 

\emailAdd{skravec@ucsd.edu}
\emailAdd{srpal@ucsd.edu}

\abstract{
We study Schr\"odinger invariant field theories (nonrelativistic conformal field theories) in the large charge (particle number) sector. We do so by constructing the effective field theory (EFT) for a Goldstone boson of the associated $U(1)$ symmetry in a harmonic potential. This EFT can be studied semi-classically in a large charge expansion. We calculate the dimensions of the lowest lying operators, as well as correlation functions of charged operators. We find universal behavior of three point function in large charge sector. We comment on potential applications to fermions at unitarity and critical anyon systems.}
\maketitle

\section{Introduction and Summary}

Symmetry has always been a guiding principle in characterizing physical systems. While weakly coupled field theories are known to be tractable in terms of perturbation theory in coupling, often the strongly coupled ones can only be constrained by symmetry arguments. For example, the physics of low-energy quantum chromo dynamics (QCD) is captured by an effective theory of pions, whose low-energy interactions are fixed by the broken chiral symmetry.  

Conformal field theories (CFTs) are especially beautiful examples of how one can leverage the symmetry group. While generically strongly coupled, conformal symmetry almost completely fixes the behavior of correlation functions and gives non-trivial insights into the structure of their Hilbert spaces. In some cases, the conformal bootstrap \cite{Rattazzi:2008pe}
can provide us with rich physics of such theories entirely based on symmetry principles. However, we are still lacking many concrete calculational tools for these theories. In CFTs with an additional global $U(1)$, recent progress has been made by constructing effective field theories for their large charge ($Q$) sector. Generically, the large charge sector can be horribly complicated in terms of elementary fields and their interactions, but one can set up a systematic $1/Q$ expansion to probe this strongly coupled regime. This has been useful in finding the scaling of operator dimensions, and many other meaningful physical quantities \cite{hellerman2015cft, hellerman2017note,
monin2017semiclassics,banerjee2018conformal,de2018large,Mukhametzhanov:2018zja}. 

In this work, we will be dealing with systems with non relativistic scale and conformal invariance i.e. systems invariant under Schr\"odinger symmetry. While in CFT, one needs to have a external global symmetry to talk about large charge expansion, the nonrelativistic conformal field theories (NRCFTs) come with a ``natural" $U(1)$, the particle number symmetry. The Sch\"odinger symmetry group and its physical consequences have been studied in \cite{Mehen:1999nd,Nishida:2010tm,Nishida:2007pj,goldberger2015ope,Golkar:2014mwa,Pal:2018idc}. The physical importance of Schr\"odinger symmetry lies in varied realisation of the symmetry group, starting from fermions at unitarity\cite{Regal:2004zza,Zwierlein:2004zz} to  examples including spin chain models \cite{Chen:2017tij}, systems consisting of deuterons \cite{Kaplan:1998tg,Kaplan:1998we}, ${}^{133}Cs$\cite{Chin:2001uan}, ${}^{85}Rb$ \cite{Roberts:1998zz},${}^{39}K$ \cite{loftus2002resonant}. 

Such theories, similar to CFTs, admit a state-operator correspondence\cite{nishida2007nonrelativistic, goldberger2015ope} in which the dimensions of operators correspond to energy of a state in a harmonic potential\footnote{This state-operator map is different from the one discussed in \cite{Pal:2018idc} to explore the neutral sector. In  \cite{Pal:2018idc}, the map is more akin to the $(0+1)$ dimensional CFT.}. Specifically, the scaling generator $D$, which scales $\vec{x}\mapsto\lambda x$ and $t\mapsto\lambda^2 t$ for $\lambda \in \mathbf{R}$ gets mapped to the Hamiltonian ($H_\omega$) in the harmonic trap i.e. $H_\omega \equiv H+ \omega^2 C$ where $C=\frac{1}{2}\int d^dx ~ x^2 n(x)$ is the special conformal generator and $n(x)$ is the number density and $H$ is the time translation generator of the Schr\"odinger group. The parameter $\omega$ determines the strength of the potential and plays an analogous role to the radius of the sphere in the relativistic state-operator correspondence.\footnote{Here and also subsequently, we will be working in non-relativistic ``natural" units of $m=\hbar=1$}. 

Given this set up, we consider an operator $\Phi$ with large number charge $Q$. For example, one can think of  $\phi^\frac{N}{2}$ for $\phi(x)=\ \normord{\psi^\dag_{\uparrow}(x) \psi^\dag_{\downarrow}(x)}$ in the case of fermions at unitarity in $d=3$ dimensions. By the state-operator correspondence, the operator is related to a state $|\Phi\rangle$ with finite density of charge ($n$) in the harmonic trap. There's an energy scale set by the density $\Lambda_{UV} \sim \mu \sim n^{\frac{2}{d}}$, $\mu$ being the chemical potential which fixes the total charge to $Q$. There is also a scale set by the trap $\Lambda_{IR} \sim \omega$ which controls the level spacing of $H_\omega$. The limit of large charge $Q \gg 1$ then implies a parametric separation of these scales. This allows us to set up a perturbatively controlled expansion in $1/Q$ and probe the large charge sector of a theory invariant under Schr\"odinger symmetry.  

In this limit it becomes appropriate to ask, what state of \emph{matter} describes the large charge sector? Such a state with finite density of charge necessarily breaks some of the space-time symmetries e.g. scale transformations, (Galilean) boosts, special-conformal transformations. That these symmetries are spontaneously broken also implies that they must be realized non-linearly in the effective field theory (EFT) describing the large charge sector. We expect the low-energy degrees of freedom to be Goldstones. 

One possibility is that the $U(1)$ symmetry remains unbroken. This is the case for a system with a Fermi surface. There the low-energy degrees of freedom would also include fermionic matter in addition to any Goldstones. The simplest candidate EFT, Landau Fermi-Liquid theory, is incompatible with the non-linearly realized Schr\"odinger symmetry\cite{rothstein2018symmetry} and therefore this is a fairly exotic possibility.

Another possibility is that the $U(1)$ symmetry is also spontaneously broken, leading to superfluid behavior. This has been the case most studied in the literature and seems like the most obvious possibility for a bosonic NRCFT. Additionally, both unitary fermions and the scale invariant anyon gas at large density are suspected to be superfluids. Therefore we focus exclusively on this symmetry breaking pattern.

\subsection*{Summary of Results}

We compute the properties of the ground state $\ket{\Phi}$ with finite density of charge, under the assumption it describes a rotationally invariant superfluid, via an explicit path integral representation:
\be \label{pathint}
\vev{\Phi|e^{-H_\omega T}|\Phi} = \int \mathcal{D}\chi ~  e^{-S_{eff}[\chi]+\mu \int d^dx ~n(x)}
\ee
where $\chi$ is a Goldstone boson describing excitations above the ground state, $\mu$ is the chemical potential and $n(x)$ is the number density which is canonically conjugate to $\chi$. This integral can then be computed by saddle point in the large $\mu$ limit. The chemical potential $\mu$ can then be fixed semi-classically in terms of the charge $Q$. Thus self-consistently, we are obtaining a large $Q$ expansion. We employ the coset construction to write down the most general effective action for the Goldstone which is consistent with the non-linearly realized Schr\"odinger symmetry. 

\begin{itemize}
\item 
For the case with magnetic vector potential $\vec{A}=0$ (the one that is relevant for the NRCFT in harmonic trap), we find the effective Lagrangian given by
\be \label{secondorderLefff}
\mathcal{L}_{eff}= c_0 X^{\frac{d}{2}+1} + c_1 \frac{X^{\frac{d}{2}+1}}{X^3} \partial_i X \partial^i X + c_2 \frac{X^{\frac{d}{2}+1}}{X^3} (\partial_i A_0)^2 + c_3 \frac{X^{\frac{d}{2}+1}}{X^2} \partial_i \partial^i A_0 + c_4 \frac{X^{\frac{d}{2}+1}}{X^2} (\partial_i \partial^i \chi)^2
\ee

where $X=\partial_t \chi-A_0 -\frac{1}{2}\partial_i \chi \partial^i \chi $. However this is not the full set of constraints. It can be shown that imposing `general coordinate invariance' will reduce the number of independent Wilson coefficients even further\cite{son2006general}. In particular there are the additional constraints: $c_2 = 0$ and $c_3 = - d^2 c_4$. 
Additionally, in $d=2$, one can have parity violating operator at this order: 
\be 
c_5 \frac{1}{X} \epsilon^{ij} (\partial_i A_0)(\partial_j X)
\ee
The details can be found in Section \ref{sec:effL}.
\item The dispersion relation of low energy excitation above the ground state is found out to be:
\be 
\epsilon(n,\ell) = \pm \omega \left(\frac{4}{d}n^2 + 4n + \frac{4}{d}\ell n - \frac{4}{d}n +  \ell \right)^{\frac{1}{2}}
\ee
where $\ell$ is the angular momentum and $n$ is a non-negative integer and $\epsilon(n,\ell)$ is the excitation energy. The dispersion determines the low-lying operator dimensions explicitly. Since, $\epsilon(n=0,\ell=1) = \pm \omega$ and $\epsilon(n=1,\ell=0)=\pm 2\omega$, they can be identified with two different kinds of descendant operators appearing in the Schr\"odinger algebra. The details can be found in Section \ref{sec:exc}.
\item 
In the leading order in $Q$, we find the ground state energy i.e. dimension $\Delta_Q$ of the corresponding operator $\Phi$:
\begin{align}
\Delta_Q=\left(\frac{d}{d+1}\right)\xi Q^{1+\tfrac{1}{d}}\,,\quad\text{where}\quad \frac{1}{c_0}=\frac{\Gamma(\tfrac{d}{2}+2)}{\Gamma(d+1)}(2\pi \xi^2)^{\tfrac{d}{2}}\,.
\end{align}
where $c_0$ is UV parameter of the theory, appearing in the Lagrangian \eqref{secondorderLefff}. 

 Specifically, we have
\begin{align}
\Delta_Q&=\frac{2}{3} \left(\xi Q^{3/2}\right) + c_1 \frac{4\pi}{3}\xi \left( Q^{\frac{1}{2}} \log Q \right)+\mathcal{O}\left(Q^{\frac{1}{2}} \right) \quad \text{for}\ d=2\,.\\
\Delta_Q&= \left(\frac{3}{4}\right)\xi Q^{4/3}-\left (c_1+\frac{c_3}{2}\right) (3\sqrt{2}\pi^2) \xi^2 Q^{2/3}+ O\left(Q^{5/9}\right)\, \quad \text{for}\ d=3\,.
\end{align}
The details can be found in Section \ref{sec:dim}.
\item We find the structure function $F$ appearing in three point function of two operators with large charge $Q$ and $Q+q$ and one operator $\phi_q$ with small charge $q$ goes as follows: 
\be \label{Ffunction}
F(v= \ii \omega y^2) \propto Q^{\frac{\Delta_\phi}{2d}} \left(1- \frac{\omega y^2 }{2\xi }Q^{-1/d}\right)^{\frac{\Delta_\phi}{2}} e^{-\frac{1}{2} q \omega y^2}
\ee
where $y$ is the insertion point of $\phi_q$ in the oscillator co-ordinate and $\Delta_\phi$ is the dimension $\phi_q$. The details can be found in Section \ref{sec:3pt}.
\end{itemize}

\textit{Note: While this work was being completed a paper appeared with some overlap\cite{Favrod:2018xov}. They identify many of the same operators we do, through different means and without couplings to the background gauge field. The primary tool we utilize is the state-operator correspondence for NRCFTs, therefore directly compute properties of the NRCFTs in harmonic trap in large charge limit.}

\section{Lightning Review of Schr\"odinger Algebra}

The Schr\"odinger algebra has been extensively explored in \cite{Mehen:1999nd,Nishida:2010tm,Nishida:2007pj,goldberger2015ope,Golkar:2014mwa,Pal:2018idc}. Here we take the readers through a quick tour of the essential features of Schr\"odinger algebra, that we are going to use through out this paper. The most important subgroup of Schr\"odinger group is the Galilean group, generated by time translation generator $H$, spatial translation generators $P_i$, rotation generators $J_{ij}$ and boost generators $K_i$. One can centrally extend this group by appending another $U(1)$ generator $N$, which generates the particle number symmetry. As a whole, these generators constitute what we call Galilean algebra and they satisfy:
\begin{gather}
[J_{ij},N]=[P_i,N]=[K_i,N]=[H,N]=0\nonumber \\
 [J_{ij}, P_{k}]=\ii
(\delta_{ik}P_{j}-\delta_{jk}P_{i})\,, \nonumber \\
 [J_{ij},K_{k}]=\ii
(\delta_{ik}K_{j}-\delta_{jk}K_{i})\,, \nonumber\\
[J_{ij},J_{kl}]= \ii (\delta_{ik}J_{jl}-\delta_{jk}J_{il}+\delta_{il}J_{kj}-\delta_{jl}J_{ki})\,,\nonumber \\
\label{eq:GalAlg1} 
[P_{i},P_{j}]=[K_{i},K_{j}]=0\,, \qquad [K_{i},P_{j}]=\ii\delta_{ij}N\,,\\
[H,N]=[H,P_{i}]=[H,M_{ij}]=0\,, \quad [H,K_{i}]=-\ii P_{i}\,. \nonumber  
\end{gather}

The Galilean group is enhanced to Schr\"odinger group by appending a scaling generator $D$ and a special conformal generator $C$ such that they satisfy the following commutator relations: 
\begin{gather}
[D,P_{i}]=\ii P_{i}\,, \quad [D,K_{i}]= -\ii K_{i}\,,  \\
\quad [D,H]=2\ii H\,,\quad [D,C]=-2\ii C\,, [H,C]=-\ii D\,,\\
\quad [J_{ij},D]=0\,,\quad [J_{ij},C]=0\,,\quad [N,D]=[N,C]=0\,.
\end{gather}

The state-operator correspondence for an NRCFT is based on the following definition \cite{goldberger2015ope}:
\be \label{SOdef}
\ket{\mathcal{O}} \equiv e^{-\frac{H}{\omega}} \mathcal{O}^\dag (0) \ket{0} = \mathcal{O}^\dag\left(-\frac{\ii}{\omega},0\right)\ket{0}
\ee
where $\mathcal{O}^\dag$ is a primary operator of number charge $Q_{\mathcal{O}^\dag} = -Q_{\mathcal{O}} \geq 0$. By the Schr\"odinger algebra, this state satisfies:
\be \label{SOconseq}
N \ket{\mathcal{O}} = Q_{\mathcal{O}^\dag}\ket{\mathcal{O}} ~~~~~ H_\omega \ket{\mathcal{O}} = \omega \Delta_\mathcal{O} \ket{\mathcal{O}}
\ee
where $H_\omega = H + \omega^2 C$ is the Hamiltonian with the trapping potential.

It is natural to define a transformation from Galilean coordinates $x=(t,\vec{x})$ to the ``oscillator frame" $y=(\tau, \vec{y})$ where the time translation $\tau \rightarrow \tau + a$ is generated by $H_\omega$. Explicitly this is given by
\be \label{coordinates}
\omega \tau = \arctan \omega t\,,\quad\quad  \vec{y} = \frac{\vec{x}}{\sqrt{1+\omega^2 t^2}}
\ee
and allows us to map primary operators and their correlation functions in the oscillator frame to the Galilean frame via the map\cite{goldberger2015ope}:
\begin{align}\label{primopmap}
\mathcal{\tilde{O}}(y) &= \left(1+\omega^2 t^2\right)^{\frac{\Delta_\mathcal{O}}{2}} \exp\left[\frac{\ii}{2} Q_{\mathcal{O}} \frac{ \omega^2 |\vec{x}|^2 t}{1+\omega^2 t^2}\right] \mathcal{O}(x)\\
\mathcal{O}(x)&=\left[\cos(\omega t)\right]^{\Delta_{\mathcal{O}}} \exp\left[-\frac{\ii}{2} Q_{\mathcal{O}} \omega |\vec{y}|^2 \tan(\omega\tau)\right] \mathcal{\tilde{O}}(y)
\end{align}

In this paper, we will be interested in matrix elements of the form:
\be \label{genericC}
\vev{\Phi|\phi_1(y_1) \cdots \phi_n(y_n)|\Phi}
\ee
where $\Phi^\dag$ is a primary of charge $Q \gg 1$ and $\phi_i$ are also charged \footnote{The state-operator correspondence breaks down for neutral operators as they actually trivially on the vacuum and their representation theory is not well understood. \cite{Pal:2018idc} explores how to circumvent this issue.} primaries with $q_i \ll Q$.\footnote{Here we point out that if an operator is explicitly written as a function of oscillator co-ordinate, it is to be understood that we have already employed the mapping \eqref{primopmap}. Thus $\phi_{i}(y_1)$ in \eqref{genericC} should technically be written as $\tilde{\phi}_{i}(y_1)$, albeit we omit ``tilde" sign for notational simplicity.}

In the Galilean frame, the general form of a two point function is fixed to be
\begin{align}
\vev{\mathcal{O}_1(x_1)\mathcal{O}_2(x_2)} = c \delta_{\Delta_1,\Delta_2} \delta_{Q_1,-Q_2} \frac{\exp\left[\ii Q_{2}\frac{|\vec{x}|^2}{2t}\right]}{(t_1-t_2)^{\Delta_1}}
\end{align}
where $c$ is a numerical constant, $\Delta_i$ is the dimension of the operator $\mathcal{O}_i$, $Q_i$ is the charge of $\mathcal{O}_i$. The symmetry algebra constrains the general form of a three-point function upto a arbitrary function of a cross-ratio $v_{ijk}$ defined below:
\begin{align} \label{3ptfuncGal}
\nonumber &\vev{\mathcal{O}_1(x_1)\mathcal{O}_2(x_2)\mathcal{O}_3(x_3)}\equiv G(x_1;x_2;x_3)\\
&= F(v_{123}) \exp\left[-\ii \frac{Q_1}{2} \frac{\vec{x}_{13}^2}{t_{13}}-\ii \frac{Q_2}{2} \frac{\vec{x}_{23}^2}{t_{23}}\right] \prod_{i<j} t_{ij}^{\frac{\Delta}{2}-\Delta_i -\Delta_j} 
\end{align}
where $\Delta\equiv \sum_i \Delta_i $ , $x_{ij}\equiv x_i - x_j$ , and $F(v_{ijk})$ is a function of the cross-ratio $v_{ijk}$ defined:
\be \label{crossratio}
v_{ijk} = \frac{1}{2}\left(\frac{\vec{x}_{jk}^2}{t_{jk}} - \frac{\vec{x}_{ik}^2}{t_{ik}} + \frac{\vec{x}_{ij}^2}{t_{ij}} \right)
\ee
We note that the three point function becomes zero unless $\sum Q_i=0$.

\section{Lightning Review of Coset Construction}

A symmetry is said to be spontaneously broken if the lowest energy state, the ground state, is not an eigenstate of the associated charge. The low-energy effective action, describing the physics above the ground state, is still invariant under the full global symmetry group but the broken subgroup is realized \emph{non-linearly}. Typically this means the effective action describes some number of Goldstones.

The coset construction gives a general method for constructing effective actions with appropriate non-linearly realized symmetry actions. It was developed for internal symmetries by CCZW \cite{coleman177structure, callan1969cg} and later generalized to space-time symmetries\cite{ogievetsky1974nonlinear}. Here we give a nimble review of the method and its application to the superfluid. We refer to the original literature and the recent review \cite{delacretaz2014re} for more details. The primary objective of the coset construction is to write down the most general action, invariant under a global symmetry group $G$ but where only the subgroup $G_0$ is linearly realized. Let us consider a symmetry group which contains the group of translations, generated by $P_a$. Let us denote the broken generators as $X_b$ corresponding to associated Goldstones $\pi_b(x)$. We denote unbroken generators as $T_c$.

We can define the exponential map from space-time to the coset space $G / G_0$
\be \label{expmap}
U \equiv e^{\ii \bar{P}_a x^a} e^{\ii X_b \pi^b(x)}
\ee
With this map we can define the $1$-form, known as the Maurer-Cartan (henceforth we call it MC) form, on the coset space. Under a $G$-transformation \eqref{expmap} transforms as
\be \label{exptrans}
g: U(x) \rightarrow e^{\ii \bar{P}_a (x')^a} e^{\ii X_b \pi^{'b}(x')} h(\pi(x),g)
\ee
where $h(\pi(x),g)$ is some element in $G_0$, determined by the Goldstones and $g \in G$, that ``compensates" to bring $U(x)$ back to the form in \eqref{expmap}. This determines how the Goldstone fields transform\footnote{For space-time symmetries there's a translation piece even though $\bar{P}_a$ are unbroken. This is because, on coordinates, translations are always non-linearly realized as $x \rightarrow (x+a)$}.

Expanded in a basis of generators the MC form looks like:
\be \label{MCformgen}
\Omega \equiv -\ii U^{-1} \partial_\mu U \equiv E_\mu^a(\bar{P}_a + (\nabla_a \pi^b) X_b + A_a^c T_c )
\ee
where each of the tensors $\{E_\mu^a , ~ \nabla_a \pi^b , ~ T_c \}$ is a function of the Goldstone fields $\pi_a$. Here $E_\mu^a$ is a vierbein, $\nabla_a \pi^b$ are the covariant Goldstone derivatives and $A_a^c$ transforms like a connection. 

Several remarks are in order. Once space-time symmetries are broken the quantity $d^dx$ is no longer necessarily a scalar under those transformations. However the quantity $d^dx \det E$  can be used to define an invariant measure for the action. On the other hand, contractions of the objects $\nabla_a \pi^b$, in a way which manifestly preserves the $G_0$ symmetry, also provides us with $G$ invariants and form the Goldstone part of the effective action. The connection, $A_a^c$ and the vierbein, can be used to define the following ``higher" covariant derivative
\be \label{highcov}
\nabla^H_a \equiv (E^{-1})_a^\mu \partial_\mu + \ii A_a^c T_c
\ee
An object like $\nabla^H_a \nabla_b \pi^c$ also transforms covariantly and $G_0$-invariant contractions with other tensors should be included. The other primary use of \eqref{highcov} is for defining covariant derivatives of ``matter fields". For example, suppose $\psi$ is a matter field transforming in a $k$-dimensional linear representation $r$ of $G_0$ as $\psi \rightarrow \psi' = r(h) \psi$. The coset construction provides multiple ways to uplift $G_0$ representations to full $G$ representations. The one of importance to us is when $r$ appears in the decomposition of a $K$-dimensional representation $R$ of $G$. Defining the field $\tilde{\psi} \equiv \left( \psi , ~ 0 \right)$ in the $K$-dimensional representation, one can show that the field $\Psi = R(\Omega) \tilde{\psi}$ transforms linearly under the full group $G$. If a subset of the symmetry is gauged then we just covariantly replace $\partial_\mu \rightarrow D_\mu = \partial_\mu + \ii \bar{A}_\mu^d \bar{T}_d$ in the above. The tensors will then depend on the gauge fields $\bar{A}$ but otherwise everything goes through.

One last important aspect of space-time symmetry breaking is that not all the Goldstone bosons are necessarily independent \cite{low2002spontaneously}. This occurs when the associated currents differ only by functions of spacetime. A localized Goldstone particle is made by a current times a function of spacetime, so we can not sharply distinguish the resulting particles. This redundancy also appears in the coset construction. Suppose $X$ and $X'$ are two different broken generators in different $G_0$-multiplets and we denote their associated Goldstone bosons $\pi$ and $\pi'$. Let $\bar{P}_\nu$ be an unbroken translation generator. Let us also assume that there's a non-trivial commutator of the form $[P_\nu , X] \supseteq X'$. One can see, from calculating the Maurer-Cartan form via the BCH identity, that this implies an undifferentiated $\pi$ in the covariant Goldstone derivative $\nabla_\nu \pi'$. The quadratic term is then $ (\nabla_\nu \pi')^2 \sim c^2 \pi^2 $ ; this is an effective mass term for the $\pi$ Goldstone. Thus we are justified in integrating it out by imposing its equation of motion. A simpler, but equivalent up to redefinitions, constraint is setting $\nabla_\nu \pi' = 0$. This is a covariant constraint, completely consistent with the symmetries. In the literature it is known as an ``inverse Higgs constraint" .

\section{Schr\"odinger Superfluid from Coset Construction}\label{sec:effL}
In this section, we will use the coset construction to construct the most general Goldstone action consistent with the broken symmetries of a rotationally invariant Schr\"odinger superfluid. For the purpose of determining local properties of the superfluid state in the trap we can first work in the thermodynamic limit defined by $\Lambda_{IR} \sim \omega \rightarrow 0$. The symmetry generators are then just those of the usual Schr\"odinger group. 

The superfluid ground state $\ket{\Phi}$ spontaneously breaks the number charge $N$. As mentioned in the introduction, this state also breaks the conformal generators and boosts. It is simplest to describe such states in the grand canonical ensemble. We remark that in the thermodynamic limit, one can leverage the equivalence between canonical ensemble with fixed chrage and grand canonical ensemble\footnote{As a result, one can always view the large charge expansion as a large chemical potential expansion}. Thus, in what follows, we define the operator $\bar{H} = H-\mu N$ such that $\bar{H} \ket{\Phi} = 0$. The parameter $\mu$ plays the role of a chemical potential; it is a Lagrange multiplier to be determined by the charge density. By assumption, $\ket{\Phi}$ is not an eigenstate of $N$. It therefore cannot be an eigenstate of $H$ while satisfying $\bar{H}\ket{\Phi}=0$. The unbroken `time' translations are therefore generated by $\bar{H}$\cite{nicolis2013more}. The symmetry breaking pattern is then given by:
\be \label{symmetry}
\text{Unbroken:}~ \{ \bar{H} \equiv H - \mu N\,, P_i\,, J_{ij}\} ~~~~~ \text{Broken:}~ \{ N\,, K_i\,, C\,, D\}\,,
\ee
for which we can parameterize the coset space as:
\be \label{cosetelem}
U= e^{\ii \bar{H} t} e^{-\ii \vec{P}\cdot \vec{x}} e^{\ii \vec{\eta} \cdot \vec{K}} e^{-\ii \lambda C} e^{-\ii \sigma D} e^{\ii \pi N} = e^{\ii H t} e^{-\ii \vec{P}\cdot \vec{x}} e^{\ii \vec{\eta} \cdot \vec{K}} e^{-\ii \lambda C} e^{-\ii \sigma D} e^{\ii \chi N}\,.
\ee
Here we use 4 distinct Goldstone fields:
\begin{itemize}
\item $\pi$ is the `phonon', the Goldstone for the charge. It defines the shifted field $\chi \equiv \pi + \mu t$

\item $\vec{\eta}$ is the `framon', the Goldstone for (Galilean) boosts. It transforms as a vector.

\item $\lambda$ is the `trapon', the Goldstone for special conformal transformations.

\item $\sigma$ is the `dilaton', the Goldstone for dilations. 

\end{itemize}
To allow for a background field $A_\mu$, we define the covariant derivative $D_\mu = \partial_\mu + \ii A_\mu N$. From this group element we can calculate the MC form:
\be \label{MCform}
-\ii U^{-1} D_\mu U \equiv E_\mu^\nu[ \bar{P}_\nu + (\nabla_\nu \eta^i )K_i - (\nabla_\nu \lambda ) C - (\nabla_\nu \sigma) D + (\nabla_\nu \pi) Q]
\ee
where $\bar{P}_\mu \equiv (-\bar{H}, \vec{P})$, and we've anticipated the absence of a gauge field for $J_{ij}$. We remark that the relativistic notation is just for ease of writing; because space and time are treated differently we have to treat those components of the MC form separately. Explicitly we have the following:
\be \label{vierbein}
E_0^0 = e^{-2\sigma}\,, ~~~~~ E_0^i = -\eta^i e^{-\sigma}\,,  ~~~~~ E_i^0 = 0\,,  ~~~~~ E_i^j = \delta_i^j e^{-\sigma}\,, 
\ee
\be \label{framon}
\nabla_0 \eta^j =  e^{3 \sigma}( \dot{\eta}^j + \vec{\eta}\cdot \vec{\partial}\eta^j )\,,  ~~~~~ \nabla_i \eta^j = e^{2 \sigma}( \partial_i \eta^j - \lambda \delta_i^j )\,, 
\ee
\be \label{trapon}
\nabla_0 \lambda = e^{4\sigma}(\dot{\lambda}+ \vec{\eta}\cdot \vec{\partial}\lambda +  \lambda^2)\,,  ~~~~~ \nabla_i \lambda = e^{3\sigma}\partial_i \lambda\,, 
\ee
\be \label{dilaton}
\nabla_0 \sigma = e^{2\sigma}(\dot{\sigma}+\vec{\eta}\cdot \vec{\partial}\sigma - \lambda)\,,  ~~~~~ \nabla_i \sigma = e^\sigma \partial_i \sigma\,, 
\ee
\be \label{phonon}
{\nabla_0 \pi} = e^{2\sigma}(\dot{\chi} - A_0 -\mu e^{-2\sigma}+ \vec{\eta}\cdot \vec{\partial}\chi  + \frac{1}{2}\eta^2)\,,  ~~~~~ \nabla_i \pi = e^\sigma ( \partial_i \chi - A_i + \eta_i)\,, 
\ee
which can be used to construct the effective action.

There are 4 commutators that each imply a different constraint
\be \label{piIHC}
[P_i , K_j] = - \ii \delta_{ij} N \implies \nabla_i \pi = 0\,,  ~~~~~~ [\bar{H} , D ] = -2\ii (\bar{H} + \mu N) \implies \nabla_0 \pi = 0\,, 
\ee
\be \label{confIHC}
[\bar{H}, C ] = -\ii D \implies \nabla_0 \sigma = 0\,,  ~~~~~~ [P_i , C] = -\ii K_j \delta_{ij} \implies \nabla_i \eta^j = 0\,.
\ee
Imposing them allows everything to be written in terms of a single Goldstone field $\chi$. Upon defining the gauge invariant derivatives:
\be \label{gaugeinv}
D_t \chi \equiv \partial_t \chi - A_0\,, ~~~~ D_i \chi \equiv \partial_i \chi - A_i\,,
\ee
the simplest pair can be solved as:
\begin{eqnarray}\label{superfluid}
\nabla_i \pi = 0 &\implies &  \eta_i = -D_i \chi\,,\\
\nabla_0 \pi = 0 &\implies & \mu e^{-2 \sigma} = D_t \chi - \frac{1}{2}D_i \chi D^i \chi \,.
\end{eqnarray}

The other two involve the trapon $\lambda$:
\begin{eqnarray}\label{confworked1}
\nabla_i \eta^j = 0 &\implies &   \lambda \delta_i^j= \partial_i \eta^j= - \partial_i D^j \chi\,, \\\label{confworked2}
\nabla_0 \sigma = 0 &\implies&  \lambda = \dot{\sigma}+ \vec{\eta}\cdot \vec{\partial}\sigma\,,
\end{eqnarray}
which can be written together as:
\be \label{confcont}
\dot{\sigma}+\vec{\eta}\cdot \vec{\partial}\sigma - \frac{1}{d}\vec{\partial}\cdot \vec{\eta} = -\frac{1}{2} \frac{\partial_0 X}{X} + \frac{1}{2} \frac{D_i \chi \partial^i X}{X} +  \frac{1}{d}  \partial_i D^i \chi = 0\,.
\ee
This is simply the leading order equation of motion for $\chi$ as we will show below.

The leading order action comes from the vierbein \eqref{vierbein} which can be expressed with $\chi$ as
\be \label{expvierbein}
\det E = e^{-(d+2)\sigma} \propto \left(D_t \chi-\frac{1}{2}D_i \chi D^i \chi\right)^{\frac{d}{2}+1}\,.
\ee
Defining the variable $X$ as
\be \label{Xdef}
X = D_t \chi -\frac{1}{2}D_i \chi D^i \chi\,,
\ee
we can write the leading order effective action as
\be \label{leading order}
S_0 = \int dt d^dx ~c_0\ \mathcal{O}_0=\int dt d^dx ~ c_0 X^{\frac{d}{2}+1}\,,
\ee
where $c_0$ is a dimensionless constant. The leading order theory \eqref{leading order} is time reversal invariant as it acts as:
\be \label{T reversal}
T: ~~~ t\rightarrow -t\,, ~~~ \pi \rightarrow - \pi \,,~~~ A_0 \rightarrow - A_0\,.
\ee
 
Higher derivative terms are constructable from contractions of the following objects:
\be \label{highdir}
\nabla_0 \eta^i\,, ~~~~ \nabla_0 \lambda\,, ~~~~ \nabla_i \lambda\,, ~~~~ \nabla_i \sigma\,. 
\ee
as well as contractions of the `higher covariants' 
\be \label{highercov}
\nabla^H_0 = -e^{2\sigma} \partial_0 + e^\sigma \eta^i \partial_i\,,  ~~~~~ \nabla^H_i = e^\sigma \partial_i\,, 
\ee
acting on the tensors \eqref{highdir}. All of these objects can be expressed in terms of $\chi$ by the constraints \eqref{piIHC} and \eqref{confIHC}. Even though we are interested in large $Q$ expansion eventually, to touch the base with the EFT written in \cite{son2006general}, we emphasize that the power counting is done with $X$, being taken to be $\mathcal{O}(p^0)$, which implies that objects like $[(\partial_i \chi)(\partial_i \chi)]^k$, $\partial_t\chi$ and $A_0$ are also order one. Additional derivatives then increase the dimension. In what follows, the field strengths$E_i$ and $F_{ij}$ are defined as
\be \label{EandBd}
E_i \equiv \partial_0 A_i - \partial_i A_0 ~~~~ F_{ij} \equiv \partial_i A_j - \partial_j A_i\,.
\ee

At $\mathcal{O}(p^2)$ we have following operators:
\begin{align} \label{gradsig}
\mathcal{O}_1 &  \equiv \det E\ \nabla_i \sigma \nabla^i \sigma\ \propto\ \frac{X^{\frac{d}{2}+1}}{X^3} \partial_i X \partial^i X\,, \\
 \label{quadratic}
\mathcal{O}_2 &  \equiv \det E\ (\nabla_0 \eta_i - 2 \nabla_i \sigma)^2\ \propto\ \frac{X^{\frac{d}{2}+1}}{X^3}[E^2 + 2 E_i F_{ij} (D_j \chi) + F_{ij} F_{ik} (D_j \chi) (D_k \chi) ]\,, \\
 \label{derivE}
\mathcal{O}_3  & \equiv \det E\ \nabla_i \sigma(\nabla_0 \eta^i - 2 \nabla^i \sigma)\ \propto\ \frac{X^{\frac{d}{2}+1}}{X^2} [\partial_i E^i + [\partial_i F_{ij} ](D_j \chi) - \frac{1}{2}F_{ij}F^{ij}]\,, \\
 \label{lambda2}
\mathcal{O}_4  & \equiv \det E\ \nabla_0 \lambda\ \propto\ \frac{X^{\frac{d}{2}+1}}{X^2} (\partial_i D^i \chi)^2\,, 
\end{align}
where the second expression of \eqref{derivE} is obtained via integration-by-parts and the \eqref{lambda2} is obtained by a straight forward application of the identity \eqref{confcont} and integration-by-parts. These operators were found in reference\cite{son2006general} for $d=3$ by very different means. Additionally, in $d=2$, one can construct following parity violating operators at this order: 
\be \label{parityvol1}
\mathcal{O}_5 \equiv \det E\ \epsilon^{ij} (\nabla_0 \eta_i ) (\nabla_j \sigma)  \propto \frac{X^{\frac{d}{2}+1}}{X^3} \epsilon^{ij} \left[E_i -F_{jk} (D_k \chi)\right](\partial_j X)\,,
\ee 
\be \label{parityvol2}
\mathcal{O}_6 \equiv \det E\ \epsilon^{ij} \nabla^H_i  (\nabla_0 \eta_j - 2 \nabla_j \sigma) \propto  \frac{X^{\frac{d}{2}+1}}{X^2} \epsilon^{ij} \partial_i (E_j-F_{jk} (D_k \chi) )\,.
\ee
Similarly in $d=3$ we have $\epsilon^{ijk}$ but that means the parity violating operators will be higher order in the derivative expansion.

\section{Superfluid Hydrodynamics}
In this section, we study the superfluid hydrodynamics. As a warm up, we first consider the fluid without the trap, thus there is no intrinsic length scale associated with such a system. The leading order superfluid Lagrangian is known to take the form \cite{son2006general}:
\be \label{pressureL}
\mathcal{L}= P(X)
\ee
where $P$ stands for `pressure' as function of the chemical potential $\mu$ at zero temperature and $X$ is the same as defined in the previous section. Due to the absence of any internal scale, dimensional analysis dictates that:
\be \label{pressurescale}
 P =c_0 \mu^{\frac{d}{2}+1}\,,
\ee
which we get from \eqref{leading order} by evaluating on the groundstate solution $\chi_{cl} = \mu t$. The number density is conjugate to the Goldstone field $\chi$ and at leading order is:
\be \label{numberdensity}
n \equiv \frac{\partial \mathcal{L}}{\partial \dot{\chi} } = P'(X)= c_0 \left(\frac{d}{2}+1\right) X^{\frac{d}{2}}\,.
\ee
One can then define the superfluid velocity in terms of the Goldstone as:
\be \label{velocity}
v_i \equiv -D_i \pi = -D_i \chi=\eta_i
\ee
where we have used the inverse Higgs constraint \eqref{superfluid}. This gives a simple interpretation of the equation of motion:
\be \label{EoMchi}
\partial_\mu \frac{\partial \mathcal{L}}{\partial (\partial_\mu \chi)} = \partial_t n + \partial_i ( n v^i) = 0\,,
\ee
which is the continuity equation of superfluid hydrodynamics. Using equations \eqref{superfluid}, we can write:
\be \label{number}
\partial_\mu n =c_0 \frac{d}{2}\left(\frac{d}{2}+1\right) X^{\frac{d}{2}-1} (\partial_\mu X )= - d n (\partial_\mu \sigma) ~~~~~ \partial_i v^i = -\partial_i D^i \chi = \vec{\partial}\cdot \vec{\eta}
\ee

The equation of motion \eqref{EoMchi} thus comes out to be as follows:
\be \label{tada}
\partial_t n + \partial_i ( n v^i) = -d n \dot{\sigma} - d n (\vec{\eta}\cdot \vec{\partial}\sigma) + n \vec{\partial}\cdot \vec{\eta}= 0
\ee
and becomes equivalent to the constraint \eqref{confcont}. Thus the superfluid EFT is consistent with the symmetry breaking pattern we discussed in the previous section.

\subsection{Superfluid in a Harmonic Trap}
Now we turn on the harmonic trap and study this superfluid EFT in the trapping potential by taking:
\be \label{trapcouple}
 A_0 = \frac{1}{2} \omega^2 r^2\,, ~~~~~ \vec{A}=0\,.
\ee
In the presence of a harmonic potential, the ground state density is no longer uniform. The number density is given by the conjugacy relation \eqref{numberdensity} and to leading order is:
\be \label{trapdensity}
n(x) = c_0 \left(\frac{d}{2}+1\right) (\mu - \frac{1}{2} \omega^2 r^2)^{\frac{d}{2}}\,,
\ee
which is vanishing at the ``cloud radius" $R =  \sqrt{\frac{2\mu}{\omega^2}}$. This defines an IR cutoff for the validity of our EFT in the trap. Semi-classically, we can fix $\mu$ in terms of the number charge $Q$ by imposing\footnote{This is equivalent to fixing $Q$ by differentiating the free energy given by the action}:
\be \label{numberfix}
Q=\vev{Q|\hat{N}|Q} = \int d^dx \vev{Q|n(x)|Q} = \frac{c_0 (2 \pi )^{d/2} \Gamma \left(\frac{d}{2}+2\right) \left(\frac{\mu }{\omega }\right)^d}{\Gamma (d+1)} \implies \frac{\mu}{\omega}\equiv \xi Q^{\frac{1}{d}}
\ee

The naive effective Lagrangian up to next-leading order is then:
\be \label{secondorderLeff}
\mathcal{L}_{eff}= c_0 X^{\frac{d}{2}+1} + c_1 \frac{X^{\frac{d}{2}+1}}{X^3} \partial_i X \partial^i X + c_2 \frac{X^{\frac{d}{2}+1}}{X^3} (\partial_i A_0)^2 + c_3 \frac{X^{\frac{d}{2}+1}}{X^2} \partial_i \partial^i A_0 + c_4 \frac{X^{\frac{d}{2}+1}}{X^2} (\partial_i \partial^i \chi)^2
\ee

For $d=2$ we have an additional parity violating operator at this order:
\be \label{specialop}
\mathcal{L}_{eff} \ni c_5 \epsilon^{ij}\frac{ (\partial_i A_0) (\partial_j X)}{X}
\ee

However, this is not the full set of constraints. It can be shown that imposing `general coordinate invariance' will reduce the number of independent Wilson coefficients even further\cite{son2006general}. In particular there are the additional constraints:
\be \label{gencorcon}
c_2 = 0 ~~~~ c_3 = - d^2 c_4
\ee
Obtaining these from the coset construction would require additionally gauging the space-time symmetries \cite{brauner2014general}. The requirement of gauging the space-time symmetries is expected as a consequence of the number operator being part of the spacetime symmetry algebra and the fact that the number symmetry has been gauged. We leave this refinement for future work. For reasons that will become clear in the next section it is not necessary to work beyond this order in the derivative expansion.

\section{Operator Dimensions}
\subsection{Ground State Energy \& Scaling of Operator Dimension}\label{sec:dim}
The ground state energy is readily computed by a Euclidean path integral, in the infinite Euclidean time separation, the path integral projects out the ground state, from which one can read off the ground state energy. A nice pedagogical example of this technique can be found in \cite{monin2017semiclassics} in context of fast spinning rigid rotor. On the other hand, from the state operator correspondence, we know that the ground state energy translated to dimension of the corresponding operator. Thus, equipped with the effective Lagrangian \eqref{secondorderLeff} obtained, the operator dimensions can be calculated via the path integral \eqref{pathint}:
\be \label{pathintdim}
\lim_{T\rightarrow \infty} \vev{Q|e^{-H_\omega T}|Q} \sim e^{-S_{eff}[\chi_{cl}]-\mu \int d^Dx ~n(x)} \sim e^{-\Delta_Q \omega T}\,,
\ee
where to leading order we have
\be \label{actioneval}
-S_{eff}[\chi_{cl}] = c_0 \Omega_d T \int_0^{R} dr~ r^{d-1} \left(\mu -\frac{1}{2}\omega ^2  r^2 \right)^{\frac{d}{2}+1} = c_0\frac{ (2\pi) ^{d/2}  \Gamma \left(\frac{d}{2}+2\right)}{\Gamma (d+2)} \left(\frac{\mu }{\omega }\right)^{d+1} \omega T\,.
\ee
Here, $\Omega_d$ is the volume factor. Combining the results of \eqref{actioneval} and \eqref{numberfix} then gives the leading order operator dimension:
\begin{align} \label{leadingopdim}
\Delta_Q = \frac{\mu}{\omega}Q-\left(-\frac{S_{eff}}{\omega T}\right)=\frac{d }{d+1}\xi Q^{1+\frac{1}{d}}\,.
\end{align}
This predicts $\Delta_Q \sim Q^{\frac{3}{2}}$ in $d=2$ and $\Delta_Q \sim Q^{\frac{4}{3}}$ in $d=3$, as in the relativistic case. That these leading order results are finite implies we can trust the EFT prediction. In general, however, the ground state energy in the trap is an infrared (IR) sensitive quantity. This becomes apparent at higher orders in the derivative expansion. 

For example, we consider the case of $d=2$. The simplest operator at next leading order is \eqref{gradsig}. To analyze its contribution, define the distance from the cloud $s$ as $r=R-s$. Its contribution to the energy, and hence the operator dimension via \eqref{actioneval}, would go like:
\be \label{divd2}
\int \text{d}^3x ~ \frac{\partial_i X \partial^i X}{X}  \sim \int_0^R \text{d}r ~ r \frac{\omega^4 r^2}{\mu - \frac{1}{2} \omega^2 r^2} \sim \mu \int \text{d}s ~ \frac{1}{s}\,,
\ee
which is log divergent for small $s$, close to the edge. For $d=3$, noticed in reference \citep{son2006general}, a divergence first appears at next-next leading order associated with the operator:
\be \label{badopd3}
\det E (\nabla_i \sigma \nabla^i \sigma)^2 \propto \frac{(\partial_i X \partial^i X)^2}{X^{\frac{7}{2}}}\,.
\ee
This leads to a power-law divergence, implying an even greater sensitivity to IR physics compared to $d=2$. Ultimately these divergences originate from the breakdown of our EFT as the superfluid gets less dense. This occurs in a small region before the edge of the cloud at radius $R^* \equiv R - \delta$ where $\delta$ is roughly the width of this region. Following \cite{son2006general}, we can estimate the size of this region as follows. One interpretation of \eqref{trapdensity} is that the chemical potential is now effectively space dependent. At the cutoff radius $R^*$, there is then an ``effective chemical potential"
\be \label{effectivepot}
\mu(r) \equiv \mu- \frac{1}{2} \omega^2 r^2\,, ~~~~~ \mu_{eff} \equiv \mu( r=R^*) = \frac{1}{2} \delta(2 R - \delta) \omega^2 \approx R \omega^2 \delta\,.
\ee
There is a length scale set by $\mu_{eff}$ which controls the EFT expansion parameter in this region. Once that length is comparable to the distance $\delta$ itself we cannot claim to control the calculation semi-classically. Using \eqref{effectivepot} this gives the estimate scaling:
\be \label{cloudedgescale}
\delta \sim \sqrt{\frac{1}{\mu_{eff}}} \implies \delta \sim \frac{1}{(\omega^2 \mu)^{\frac{1}{6}}}
\ee

We can estimate the contribution of this region to the energy by cutting off the divergent integrals at $R^*$. For $d=2$ the effective action contains a term:
\be \label{d2NLOterm}
-S_{eff} \ni c_1 (2\pi) T \int_0^{R^*} dr ~ r \frac{\omega^4 r^2}{\mu - \frac{1}{2}\omega^2 r^2} = 4\pi  T \mu  c_1 \left( \frac{13}{8}- \log\left[ \frac{2\mu}{\mu_{eff}}\right]\right)+ \cdots
\ee
where the $\cdots$ terms vanish as $\delta \rightarrow 0 $

Substituting the relations \eqref{numberfix} and \eqref{cloudedgescale} gives:
\be \label{d2NLOterm2}
\Delta_Q \ni -4\pi  \xi Q^{\frac{1}{2}} c_1 \left( \frac{13}{8} -\frac{1}{2}\log 2- \frac{1}{3}\log Q - \frac{2}{3}\log \xi\right)
\ee
Changing the cutoff relation \eqref{cloudedgescale} by a factor can then change the $\mathcal{O}(Q^{\frac{1}{2}} )$ contribution, but not the logarithmic divergence which is universal. This translates to an uncertainty of order $\mathcal{O}(Q^{\frac{1}{2}})$ in the operator dimension in $d=2$. A similar analysis\cite{son2006general} for $d=3$ and \eqref{badopd3} translates to uncertainty of order $\mathcal{O}(Q^{\frac{5}{9}})$. 

Unlike $d=2$, the operator \eqref{gradsig} gives a finite correction to leading order scaling of dimension of operator in $d=3$. This can be found by figuring out the contibution to $S_{eff}$ [see Eq.~\eqref{secondorderLeff}]
\begin{align}
-S_{eff}\ni c_1\int \text{d}\tau^E\ \int_0^{R}\ \text{d}r\ 4\pi r^2\ \left(\frac{\omega^4r^2}{\sqrt{\mu-\tfrac{1}{2}\omega^2r^2}}\right) =c_1 (3\sqrt{2}\pi^2) \left(\frac{\mu}{\omega }\right)^2\omega T
\end{align}
Similar contribution\footnote{Contribution should have come from \eqref{quadratic} as well, but as we mentioned earlier, $c_2=0$ \cite{son2006general}.} comes from \eqref{derivE}: 
\begin{align}
-S_{eff}\ni c_3\int \text{d}\tau^E\ \int_0^{R}\ \text{d}r\ 4\pi r^2(\omega^2) \left(\mu-\tfrac{1}{2}\omega^2r^2\right)^{\tfrac{1}{2}} =c_3 \left(\frac{3\pi^2}{\sqrt{2}}\right) \left(\frac{\mu}{\omega }\right)^2\omega T
\end{align}

To summarize, using \eqref{leadingopdim}, we have
\begin{align}
\label{eq:m1}
\Delta_Q&=\frac{3}{4}\left(\xi Q^{4/3}\right) - \left(c_1+\frac{c_3}{2}\right) (3\sqrt{2}\pi^2) \xi^2 Q^{2/3}+\mathcal{O}(Q^{\frac{5}{9}}) \quad \text{for}\ d=3\,,\\
\label{eq:m2}
\Delta_Q&=\frac{2}{3} \left(\xi Q^{3/2}\right) + c_1 \frac{4\pi}{3}\xi \left( Q^{\frac{1}{2}} \log Q \right)+\mathcal{O}\left(Q^{\frac{1}{2}} \right) \quad \text{for}\ d=2\,.
\end{align}
The Eq.~\eqref{leadingopdim}, \eqref{eq:m1} and \eqref{eq:m2} constitute the main findings of this subsection.

\subsection{Excited State Spectrum}\label{sec:exc}

We can also analyze the low energy excitations above the ground state. These correspond to low lying operators in the spectrum at large charge. To compute their dimension, we expand the leading action \eqref{leading order} to quadratic order in fluctuations $\pi$ about the semi-classical saddle, $\chi = \mu t + \pi$. The spectrum of $\pi$ can then be found by linearizing the equation of motion \eqref{EoMchi}:
\be \label{linearEOM}
\ddot{\pi} - \frac{2}{d}\left(\mu - \frac{1}{2} \omega^2 r^2\right)\partial^2 \pi + \omega^2 \vec{r}\cdot \vec{\partial}\pi = 0
\ee 
Expanding $\pi(t,x) = e^{\ii \epsilon t} f(r) Y_\ell$ where $Y_\ell$ is a spherical harmonic, one can show \eqref{linearEOM} reduces to a hypergeometric equation. Details can be found in Appendix A. The dispersion relation is given by:
\be \label{dispersion}
\epsilon(n,\ell) = \pm \omega \left(\frac{4}{d}n^2 + 4n + \frac{4}{d}\ell n - \frac{4}{d}n +  \ell \right)^{\frac{1}{2}}
\ee
where $\ell$ is the angular momentum and $n$ is a non-negative integer. 
In the NRCFT state-operator correspondence, there are two different operators which generate descendants. In the Galilean frame, these are the operators $\vec{P}$ and $H$. While $\vec{P}$ raises the dimension by 1 and carries angular momentum, acting by $H$ raises the dimension by 2 and carries no angular momentum. In the oscillator frame, this corresponds to:
\be \label{oscillatorraise}
\vec{P}_\pm = \frac{1}{\sqrt{2\omega}}\vec{P}\pm \ii \sqrt{\frac{\omega}{2}} \vec{K} ~~~~~ L_\pm = \frac{1}{2}(\frac{1}{\omega}H - \omega C \pm \ii D )
\ee
which then satisfy
\be \label{raiseit}
[H_\omega , \vec{P}_\pm ] = \pm \omega \vec{P}_\pm ~~~~~ [H_\omega , L_\pm ] = \pm 2 \omega L_\pm
\ee
One can check by equation \eqref{dispersion} that $\epsilon(n=0,\ell=1) = \pm \omega$ and $\epsilon(n=1,\ell=0)=\pm 2\omega$. This allows us to identify these Goldstone modes with the descendant operators in \eqref{oscillatorraise} as $\pi_{(n=0,\ell=1)} \sim P_\pm$ and $\pi_{(n=1,\ell=0)} \sim L_\pm$. The other modes generate distinct primaries and descendants, including higher spin. We remark that in a strict sense, the above is the leading order result for the difference in dimensions between low-lying operators in this sector and the dimension of the ground state found in the previous section. It is also subject to corrections suppressed in $1/Q$ from subleading operators and loop effects.

\section{Correlation Functions}
In a relativistic CFT, the form of two and three point correlators is entirely fixed by symmetry. However, the four-point function depends on two conformally invariant cross ratios of the coordinates. The Schr\"odinger symmetry is less constraining, as there exists an invariant cross ratio even for a three-point function. This implies only the two-point functions of (number) charged operators is completely determined by symmetry. 
\subsection{Two Point Function}
Following \cite{monin2017semiclassics}, we start with analyzing two point function. In path integral approach, when the in and out states are well separated in time, we have
\begin{align}
\langle \Phi_{Q},\tau_2 | e^{-H_{\omega}(\tau^{(E)}_2-\tau^{(E)}_1)}|\Phi_{Q},\tau_1\rangle = e^{-\Delta_{\mathcal{O}}(\tau^{(E)}_2-\tau^{(E)}_1)}
\end{align}
where $\tau^{(E)}$ is the Euclideanized oscillator time. This is obtained from $\tau$ by doing Wick rotation i.e. $\tau^{(E)}=\ii \tau$. This is evidently consistent with \eqref{eq:twopoint} upon doing the Wick rotation and taking $(\tau^{(E)}_2-\tau^{(E)}_1)\rightarrow\infty$. One subtle remark is in order: the Hamiltonian $H_{\omega}$ generates the time ($\tau$) translation in oscillator frame. Thus the states prepared by path integration corresponds to operators in oscillator frame. 
\subsection{Three Point Function}\label{sec:3pt}
We consider the matrix element that defines the simplest charged\footnote{The additional charge of $\bra{\Phi}$ is required for the correlator to be overall neutral and therefore non-vanishing.} three-point function
\be \label{3ptelem}
\vev{\Phi_{Q+q}|\phi_q(y)|\Phi_Q}
\ee
where $\phi_q$ is a light charged scalar primary with charge $q$ and both of $\Phi_Q$ and $\Phi_{Q+q}$ has $\mathcal{O}(1)$ dimension, given by $\Delta_Q$ and $\Delta_{Q+q}$. By assumption, $\phi_q$ transforms in a linear representation $R$ of the unbroken rotation group. To enable calculation in our EFT, we can extend this to a linear representation of the full Schroedinger group using the Goldstone fields. In what follows, we take $\phi_q$ as the ``dressed" operator\cite{monin2017semiclassics}:
\be \label{dressing}
\phi_q(y) = R\left[e^{\ii \vec{K}\cdot \vec{\eta}} e^{-\ii \lambda C} e^{-\ii \sigma D} e^{\ii \chi N} \right] \hat{\phi}_q 
\ee
where, by the assumption of $\phi_q$ being a scalar primary, is trivially acted on by $\vec{K}$ and $C$. This, combined with \eqref{superfluid} gives
\be \label{opexpand}
\phi_q = c_q X^{\frac{\Delta_\phi}{2}} e^{\ii \chi q} 
\ee
where $c_q$ is a constant, which depends on UV physics. Upon evaluating \eqref{3ptelem} semi-classically about the saddle we found before, the leading order result for the correlator comes out to be:
\begin{align} \label{semiclassic3pt}
\nonumber \vev{\Phi_{Q+q}(\tau_2)|\phi_q (\tau,\vec{y})|\Phi_Q(\tau_1)} &= c_q \left(\mu- \frac{1}{2}m \omega^2 y^2\right)^{\frac{\Delta_\phi}{2}} e^{\ii \mu q (\tau-\tau_{2})}e^{-\ii\Delta_{Q}(\tau_2-\tau_1)}\\
 &=c_q \mu^{\frac{\Delta_\phi}{2}}\left(1-\frac{y^2}{R^2}\right)^{\frac{\Delta_\phi}{2}}  e^{\mu q \tau^{(E)}} e^{\omega\left(-\Delta_{Q+q}\tau^{(E)}_2+\Delta_Q\tau^{(E)}_{1}\right)}
\end{align}
where we have used the following identity, which can be derived using the leading order operator dimension \eqref{leadingopdim} and \eqref{numberfix}:
\be \label{opdiff}
\frac{\Delta_{Q+q} - \Delta_Q}{q} = \alpha_0 \left(1+\frac{1}{d}\right) Q^{\frac{1}{d}} + \mathcal{O}\left(\frac{1}{Q}\right) \approx \frac{\partial \Delta_Q}{\partial Q} =  \frac{\mu}{\omega}
\ee
as expected since $\mu$ is a chemical potential and $\omega \Delta_Q$ is the energy. We note that the operator insertion should be away from the edge of the cloud $|y-R| \gg \delta$, where $\delta$ is the cut-off imposed to keep the divergences coming from the $y\rightarrow R$ limit at bay. 

Now we use (the details can be found in appendix~[\ref{app:2pt}])
\begin{align}
\nonumber\lim_{\tau^{(E)}_2\rightarrow \infty}\frac{1}{(1+\omega^2t_2^2)^{\Delta_{Q+q}/2}}\exp\left(-\omega\Delta_{Q+q}\tau^{(E)}_2\right) &= 2^{-\Delta_{Q+q}}\omega^{\Delta_{Q+q}/2}\,,\\
\nonumber\lim_{\tau^{(E)}_1\rightarrow -\infty}\frac{1}{(1+\omega^2t_1^2)^{\Delta_{Q}/2}}\exp\left(\omega\Delta_{Q}\tau^{(E)}_1\right) &= 2^{-\Delta_{Q}}\omega^{\Delta_{Q}/2}\,,
\end{align}
to write down the correlator in terms of operators in Galilean frame (we repeat that the path intergral in oscillator frame prepares a state corresponding to operator in oscillator frame):
\begin{align}\label{eq:3ptlc}
\vev{\Phi_{Q+q}(\ii/\omega)|\phi_q (\tau,\vec{y})|\Phi_Q(-\ii/\omega)}= c_q \mu^{\frac{\Delta_\phi}{2}}\left(1-\frac{y^2}{R^2}\right)^{\frac{\Delta_\phi}{2}}  e^{\mu q \tau^{(E)}} 2^{-\Delta_{Q}-\Delta_{Q+q}}\omega^{(\Delta_{Q}+\Delta_{Q+q})/2}\,.
\end{align}

This can be matched onto the three point function, which is constrained by Schr\"odinger algebra:  
\begin{align}\label{eq:3ptsym}
\vev{\Phi_{Q+q}|\phi_q (\tau,\vec{y})|\Phi_Q} &=F(v) \exp\left(\frac{q}{2} \omega  y^2\right)  (2)^{\Delta_{\phi}} \left(\frac{\ii\omega}{2}\right)^{\frac{\Delta}{2}}e^{- i \omega \left(\Delta_{Q}-\Delta_{Q+q}\right)\tau}.
\end{align}
The appendix~[\ref{app:3pt}] has the necessary details. Now, upon comparing \eqref{eq:3ptsym} and \eqref{eq:3ptlc}, we deduce the universal behavior of $F(v)$ in the large charge sector: 
\be \label{Ffunction}
F(v= \ii \omega y^2) \propto Q^{\frac{\Delta_\phi}{2d}} \left(1- \frac{\omega y^2 }{2\xi }Q^{-1/d}\right)^{\frac{\Delta_\phi}{2}} e^{-\frac{1}{2} q \omega y^2}
\ee
which can be rewritten as following, using \eqref{leading order}:
\begin{align} \label{Ffunction2}
F(v= \ii \omega y^2) \propto \Delta_Q^{\frac{\Delta_\phi}{2(d+1)}}\left(1- \frac{\omega y^2}{2\xi}(\tfrac{d+1}{d\xi}\Delta_Q)^{-\tfrac{1}{d+1}} \right)^{\frac{\Delta_\phi}{2}} e^{-\frac{1}{2} q \omega y^2}
\end{align}

The \eqref{Ffunction} and \eqref{Ffunction2} are the main results of this subsection. This shows the universal scaling behavior of the structure function $F$ in the large charge sector.

\section{Conclusions and Future Directions}
We have studied the large charge ($Q$) sector of theories invariant under Schr\"odinger group. We have employed coset construction to write down an effecive field theory (EFT) describing the large $Q$ sector in any arbitrary dimension $d\geq2$ assuming superfluidity and rotational invariance. The effective Lagrangian is given by
\be 
\nonumber \mathcal{L}_{eff}= c_0 X^{\frac{d}{2}+1} + c_1 \frac{X^{\frac{d}{2}+1}}{X^3} \partial_i X \partial^i X + c_2 \frac{X^{\frac{d}{2}+1}}{X^3} (\partial_i A_0)^2 + c_3 \frac{X^{\frac{d}{2}+1}}{X^2} \partial_i \partial^i A_0 + c_4 \frac{X^{\frac{d}{2}+1}}{X^2} (\partial_i \partial^i \chi)^2
\ee
where $X=\partial_t \chi-A_0 -\frac{1}{2}\partial_i \chi \partial^i \chi $ and $\chi$ is the Goldstone excitation of the superfluid ground state. We emphasize that the general co-ordinate invariance, as discussed in \cite{son2006general} will put more constraints on the Wilson coefficients, we leave that as a future project. The EFT is then studied perturbatively as an expansion in $1/Q$. This is to be contrasted with the EFT written down in \cite{son2006general}. While EFT in \cite{son2006general} is controlled by small momentum parameter, ours is controlled by $1/Q$ expansion, which enables us to probe and derive universal results and scaling behaviors in large $Q$ sector. In particular, when $Q$ is very large, we find the scaling behavior of operator dimension with charge, consistent with that found very recently in \cite{Favrod:2018xov}. We also find that in the large charge sector, structure function of three point correlator has a universal behavior. Last but not the least we derived the dispersion relation for the low energy excitation over this state with large $Q$ and identify the two different kind of descendents as two different modes of excitations. A summary of the results can be found in the introduction.

The theory of conformal, and even superconformal, anyons has been studied before in great detail \cite{doroud2018conformal, doroud2016superconformal,Nishida:2007pj,Jackiw:1991au}. In these systems there exists a simple $n$-particle operator $\mathcal{O}=(\Phi^\dag)^n$ whose dimension is given as 
\be \label{dimshortanyon}
\Delta_\mathcal{O} = n + n(n-1)\theta
\ee
where $\theta$ is the statistics parameter that arises from the Chern-Simons term of level $k$ as $\theta=\frac{1}{2k}$ for bosonic theories. For large $k$ relative to $n$, close to the bosonic limit, this is known to be the ground state in the trap. It is known as the "linear solution" in the literature due to the linear dependence on $\theta$. For the superconformal theories it is a BPS operator and the dimension \eqref{dimshortanyon} is exact. A state corresponding to such an operator is not a superfluid and our theory cannot capture the physics of the system in that regime. However, it is known there is a level crossing for smaller $k$ where the ground state corresponds to an operator whose dimension is not protected by the BPS bound. For those operators the classical dimension scales as $n^\frac{3}{2}$, in agreement with our results. We are then led to believe the effective field theory we've constructed may apply to anyon NRCFTs in that regime. 

Another family of NRCFTs can be defined by the holographic constructions of McGreevy, Balasubramanian\cite{Balasubramanian:2008dm} and Son\cite{son2008toward}. It would be interesting to study these on the gravitational side in the large charge limit, as there might exist a regime where both the EFT and gravity descriptions are valid. The analog of this for the relativistic case was carried out recently\cite{loukas2018ads}. 

One can envision to extend our results in several ways. One possible extension of these results would be to study operators with large spin as well as charge. If the superfluid EFT remains valid, for sufficiently large spin, one naively expects such operators correspond to vortex configurations in the trap. This was studied in $CFT_3$, where multiple distinct scaling regimes were shown to exist \cite{Cuomo:2017vzg}. Moreover, one can generalize these results to NRCFTs with a larger internal global symmetry group or study systems where the symmetry breaking pattern is different. Potentially interesting examples include ``chiral" superfluids \cite{hoyos2014effective}, where the rotational symmetry is additionally broken by the superfluid order parameter, or the vortex lattice \cite{moroz2018effective} where the translation symmetry is spontaneously broken.
\section*{Acknowledgements}
The authors acknowledge useful comments from John McGreevy. This work was in part supported by the US Department of Energy (DOE) under cooperative research agreement DE-SC0009919. 

\appendix
\section{Appendix A: Phonons in the Trap}

We are solving equation \eqref{linearEOM} in the range of $r \in [0, R]$ where $R^2 = \frac{2 \mu}{ \omega^2}$ is the cloud radius.

Inserting $\pi \propto e^{\ii \epsilon t} f(r) Y_\ell$ and expanding in spherical coordinates: 
\be \label{linrEoM}
-\frac{\omega^2}{d}(R^2 -x^2)[\partial_r^2 f + \frac{(d-1)}{r} \partial_r f - \frac{1}{r^2}\ell (\ell+d-2) f] + \omega^2 r \partial_r f = \epsilon^2 f
\ee 
Defining the dimensionless variables $x \equiv \frac{r}{R}$ and $\lambda \equiv \frac{\epsilon}{\omega}$ and changing variables to $z=x^2$ 
\be \label{linzEoM}
-\frac{1}{d}(1-z)[4z \partial_z^2 f + 2\partial_z f + 2(d-1) \partial_z f - \frac{1}{z}\ell (\ell + d -2) f] + 2z \partial_z f = \lambda^2 f
\ee 
Equation \eqref{linzEoM} is a hypergeometric equation with two independent solutions
\be \label{pisoln1}
f(z) \sim c_1 z^{\frac{\ell}{2}}~ {}_{2} F_{1}(\alpha_{-},\alpha_{+},\gamma,z) + c_2 z^{\frac{1}{2}(2-d-\ell)} ~ {}_{2} F_{1}(\alpha',\beta',\gamma',z)
\ee
Our solution should be valid on the interval $z \in [0,1]$ where it should be regular and finite at both $z=0$ and $z=1$. Regularity at the origin kills the second solution immediately.

Therefore we have:
\be \label{pisoln2}
f(z) \sim c_1 z^{\frac{\ell}{2}}~ {}_{2} F_{1}(\alpha_{-},\alpha_{+},\gamma,z) 
\ee
where $\gamma = \ell + \frac{d}{2}$ , $\alpha_{\pm} = \frac{1}{2}(\ell + d -1) \pm \kappa$ ,  and $\kappa=\frac{1}{2}(1-2d+d^2-2\ell+\ell d +\ell^2 + d \lambda^2 )^{\frac{1}{2}}$

The function ${}_{2} F_{1}(\alpha_{-},\alpha_{+},\gamma,z) $ is finite at $z=1$ under one of the following possibilities:
\begin{enumerate}
\item The values $\alpha_{+} + \alpha_{-} < \gamma$ for any value of the arguments
\item If either $\alpha_{\pm}$ is equal to a non-positive integer
\end{enumerate}
To see this, we use the following identity and regularity of ${}_{2}F_{1}$ around $(1-z)=0$:
\begin{align} 
\nonumber {}_{2}& F_{1}(\alpha_{-},\alpha_{+},\gamma,z) =\frac{\Gamma(\gamma)\Gamma(\gamma-\alpha_+ -\alpha_-)}{\Gamma(\gamma -\alpha_-)\Gamma(\gamma-\alpha_+)}{}_{2} F_{1}(\alpha_{-},\alpha_{+},\alpha_-+\alpha_+ +1-\gamma,1-z) \\
&+\frac{\Gamma(\gamma)\Gamma(\alpha_+ +\alpha_- - \gamma)}{\Gamma(\alpha_-)\Gamma(\alpha_+)} (1-z)^{\gamma-\alpha_- -\alpha_+}{}_{2} F_{1}(\gamma-\alpha_{-},\gamma-\alpha_{+},1+\gamma-\alpha_- -\alpha_+,1-z)\\
&{}_{2} F_{1}(\alpha_{-},\alpha_{+},\gamma,z\sim 1)\sim  \frac{\Gamma(\gamma)\Gamma(\gamma-\alpha_+ -\alpha_-)}{\Gamma(\gamma -\alpha_-)\Gamma(\gamma-\alpha_+)}+\frac{\Gamma(\gamma)\Gamma(\alpha_+ +\alpha_- - \gamma)}{\Gamma(\alpha_-)\Gamma(\alpha_+)} (1-z)^{\gamma-\alpha_- -\alpha_+}
\end{align}
We can check explicitly that $\alpha_{+} + \alpha_{-} = \ell + d -1 \geq \gamma$ for $d \geq 2$, where the superfluid groundstate is possible. Therefore option (1) is ruled out.

Define $\alpha_{-} = -n$ where $n$ is a non-negative integer. 

The relation above implies $\alpha_{+} =( \ell + d - 1)-\alpha_{-}  = \ell + d + n-1$

Consider the explicit product:
\be \label{prodquad}
\alpha_{+} \alpha_{-} = \frac{1}{4}(\ell + d - 1 + 2\kappa)(\ell + d - 1 -2\kappa) = \frac{d}{4}(\ell - \lambda^2)
\ee
Substituting the integer relations for $\alpha_{\pm}$ turns equation \eqref{prodquad} into a quadratic equation which can be solved for $\lambda$ as:
\be \label{soln}
\lambda^2 = \frac{1}{d}(4n^2 + 4dn + 4\ell n - 4n + d \ell)
\ee
which yields the dispersion \eqref{dispersion}

\section{Appendix B: Correlation Functions in Oscillator Frame}
\subsection{Two point function}\label{app:2pt}
In Galilean frame the two point function is given by
\begin{align}\label{s1}
\left\langle\mathcal{O}(t=-\ii/\omega)\mathcal{O}^\dagger(t=\ii/\omega)\right\rangle = c \left(-\frac{2\ii}{\omega}\right)^{-\Delta_{\mathcal{O}}}
\end{align}

Now we know
\begin{align}\label{s2}
\nonumber \left\langle\mathcal{O}(t=-\ii/\omega)\mathcal{O}^\dagger(t=\ii/\omega)\right\rangle &= \lim_{\underset{\tau_0\rightarrow -\ii\infty}{t_0\rightarrow \ii\omega}}\frac{1}{(1+\omega^2t_0^2)^{\Delta_{\mathcal{O}}}}\left\langle \mathcal{O}(\tau=\tau_0)\mathcal{O}^\dagger(\tau=-\tau_0) \right\rangle\\
\nonumber&=c\lim_{\tau_0\rightarrow \ii\infty}\frac{1}{(1+\omega^2t_0^2)^{\Delta_{\mathcal{O}}}} \left(\frac{1}{\sin^2(2\omega\tau_0)}\right)^{\Delta_{\mathcal{O}}/2}\\
&=c  (2\ii)^{\Delta_{\mathcal{O}}}\lim_{\tau^{(E)}_0\rightarrow \infty}\frac{1}{(1+\omega^2t_0^2)^{\Delta_{\mathcal{O}}}}\exp\left(-2\omega\Delta_{\mathcal{O}}\tau^{(E)}_0\right)
\end{align}

where $\omega t_0=\tan(\omega \tau_0)$. Comparing \eqref{s1} and \eqref{s2}, we obtain an identity:
\begin{align}
\lim_{\underset{\tau^{(E)}_0\rightarrow \infty}{t_0\rightarrow \ii\omega}}\frac{1}{(1+\omega^2t_0^2)^{\Delta_{\mathcal{O}}/2}}\exp\left(-\omega\Delta_{\mathcal{O}}\tau^{(E)}_0\right) = 2^{-\Delta_{\mathcal{O}}}\omega^{\Delta_{\mathcal{O}}/2}
\end{align}
where we have $\omega t =\tan (\omega\tau)$ and $\tau^{(E)}=\ii \tau$. We note that $t=\pm \frac{\ii}{\omega}$ corresponds to Oscillator frame Euclidean time $\tau_{E}=\mp \infty$, this follows from 
\be
\omega t=\tan\left(-\ii\omega\tau_{E}\right)
\ee
Thus the operators are inserted at infinitely past and future Euclidean time.

 In the oscillator frame, we have
\begin{align}
\nonumber \vev{\mathcal{O}(\tau_1)\mathcal{O}^{\dagger}(\tau_2)}&= c \left[1+\tan^2(\omega\tau_1)\right] ^{\frac{\Delta_{\mathcal{O}}}{2}}\left[1+\tan^2(\omega\tau_2)\right]^{\frac{\Delta_{\mathcal{O}}}{2}} \left(\tan(\omega\tau_1)-\tan(\omega\tau_2)\right)^{-\Delta_{\mathcal{O}}}\,,
\end{align}
which can be simplied into
\begin{align}\label{eq:twopoint}
\vev{\mathcal{O}(\tau_1)\mathcal{O}^{\dagger}(\tau_2)}=c\left[\sin(\omega (\tau_1-\tau_2)) \right]^{-\Delta_{\mathcal{O}}}\,,
\end{align}
using the identity 
\begin{align}
\frac{[1+\tan^2(\omega\tau_1)][1+\tan^2(\omega\tau_2)]}{[\tan(\omega\tau_1)-\tan(\omega\tau_2)]^2}=\frac{1}{\sin^2(\omega(\tau_1-\tau_2))}\,.
\end{align}

\subsection{Three point function}\label{app:3pt}
In the Galilean frame, the general form of a three-point function is fixed to be:
\be \label{3ptfuncGal1}
\vev{\mathcal{O}_1(x_1)\mathcal{O}_2(x_2)\mathcal{O}_3(x_3)}\equiv G(x_1;x_2;x_3)= F(v_{123}) \exp\left[-\ii \frac{Q_1}{2} \frac{\vec{x}_{13}^2}{t_{13}}-\ii \frac{Q_2}{2} \frac{\vec{x}_{23}^2}{t_{23}}\right] \prod_{i<j} t_{ij}^{\frac{\Delta}{2}-\Delta_i -\Delta_j} 
\ee
where $\Delta\equiv \sum_i \Delta_i $ , $x_{ij}\equiv x_i - x_j$ , and $F(v_{ijk})$ is a function of the cross-ratio $v_{ijk}$ defined:
\be \label{crossratio1}
v_{ijk} = \frac{1}{2}\left(\frac{\vec{x}_{jk}^2}{t_{jk}} - \frac{\vec{x}_{ik}^2}{t_{ik}} + \frac{\vec{x}_{ij}^2}{t_{ij}} \right)
\ee
The matrix element \eqref{3ptelem} defines a 3-point function in this frame via \eqref{primopmap} and \eqref{SOdef}
\begin{align*} \label{3ptours}
&\vev{\Phi_{Q+q}|\phi_q(\tau,\vec{y})|\Phi_Q} = (1+\omega^2 t^2)^{\frac{\Delta_\phi}{2}} \exp\left[\frac{\ii}{2}q \frac{x^2 \omega^2 t}{1+\omega^2 t^2}\right] G\left(-\frac{\ii}{\omega},0~ ; t,\vec{x} ; ~\frac{\ii}{\omega},0\right)\\ 
&= F(v) (1+\omega^2 t^2)^{\frac{\Delta_\phi}{2}}  \exp\left[\frac{\ii}{2}q \frac{x^2 \omega^2 t}{1+\omega^2 t^2}\right] \exp\left[-\frac{i q x^2}{2 \left(t-\frac{i}{\omega }\right)}\right]  \prod_{i<j} t_{ij}^{\frac{\Delta}{2}-\Delta_i -\Delta_j} \\
&= F(v) \exp\left[\frac{q}{2}\frac{ \omega  x^2}{1+\omega^2 t^2}\right] (1+\omega^2 t^2)^{\frac{\Delta_\phi}{2}} \prod_{i<j} t_{ij}^{\frac{\Delta}{2}-\Delta_i -\Delta_j}\\
&=F(v) \exp\left[\frac{q}{2}\frac{ \omega  x^2}{1+\omega^2 t^2}\right]  (2)^{\frac{1}{2} (-\Delta_{Q+q}+\Delta_{\phi}-\Delta_{Q})} (\ii\omega)^{\frac{\Delta}{2}} \left(\frac{1-\ii \omega t}{1+\ii \omega t}\right)^{\frac{\Delta_{Q}-\Delta_{Q+q}}{2}}\\
&=F(v) \exp\left(\frac{q}{2} \omega  y^2\right)  (2)^{\Delta_{\phi}} \left(\frac{\ii\omega}{2}\right)^{\frac{\Delta}{2}}e^{- i \omega \left(\Delta_{Q}-\Delta_{Q+q}\right)\tau}
\end{align*}
where
\be \label{ourcross}
v =\frac{1}{2}\left(\frac{x^2 }{t-\frac{\ii}{\omega}} + \frac{x^2 }{-\frac{\ii}{\omega}-t}\right) = \frac{\ii \omega  x^2}{1+ \omega ^2 t^2}
\ee

{\bibliographystyle{bibstyle2017}

\bibliography{references}

\hypersetup{urlcolor=RoyalBlue!60!black}
}

\end{document}